\documentstyle[12pt,aasms4]{article}

\received{4 April 2000}
\accepted{24 October 2000}

\slugcomment{Accepted by PASP}

\lefthead{Kenworthy, Parry and Taylor}
\righthead{SPIRAL: a fiber-fed spectrograph for the AAT}

\begin{document}

\title{SPIRAL Phase A: A Prototype Integral Field Spectrograph\\
for the AAT.}
\author{Matthew A. Kenworthy\altaffilmark{1} Ian R. Parry\altaffilmark{2}}
\affil{Institute of Astronomy, University of Cambridge,
Madingley Road, Cambridge CB3 0HA}
\and
\author{Keith Taylor\altaffilmark{3}}
\affil{Anglo Australian Observatory, Epping, Sydney, Australia}

\altaffiltext{1}{Electronic mail: mak@as.arizona.edu}
\altaffiltext{2}{Electronic mail: irp@ast.cam.ac.uk}
\altaffiltext{3}{Electronic mail: kt@astro.caltech.edu}

\begin{abstract}

We present details of a prototype fiber feed for use on the
Anglo-Australian Telescope (AAT) that uses a dedicated fiber-fed
medium/high resolution ($R\simeq 10000$) visible-band spectrograph to
give integral field spectroscopy (IFS) of an extended object.  A focal
reducer couples light from the telescope to the close-packed lenslet
array and fiber feed, allowing the spectrograph be used on other
telescopes with the change of a single lens.  By considering the
properties of the fibers in the design of the spectrograph, an efficient
design can be realised, and we present the first scientific results of a
prototype spectrograph using a fiber feed with 37 spatial elements,
namely the detection of Lithium confirming a brown dwarf candidate and
IFS of the supernova remnant SN1987A.

\end{abstract}







\section{INTRODUCTION}

The Anglo-Australian Telescope (AAT) has pioneered many astronomical
instruments over the past 25 years, such as the 2dF galaxy survey
instrument (\cite{Tay97}) and including one of the earliest reported
fiber feeds in the literature (\cite{Gra82}).  These instruments all use
optical fibers to reformat the two dimensional distribution of light
from the focal plane of the telescope into a one dimensional fiber slit
suitable for a spectrograph, allowing multiple object (in the case of 2dF)
and spatially resolved spectroscopy of extended objects.  

Although there are many other methods used to perform integral field
spectroscopy (IFS) the extremely versatile properties of optical fibers
naturally lend themselves to optical reformatting (see \cite{Ken98} for
a review of IFS methods). 

SPIRAL (Segmented Pupil/Image Reformatting Array of Lenslets) Phase A is
an IFS design with 37 spatially resolved elements (see Figure
\ref{fig1}) that has been built to prototype the new methods needed in
building large integral field units (IFUs). Here we present a
description and the first light results of SPIRAL Phase A.

\placefigure{fig1}

The design of SPIRAL considers the optical fibers as part
of the spectrograph instead of an auxillary unit to a slit-based
spectrograph. 

In section \ref{ffs} we analyse a simple fiber-fed spectrograph.
The optical design can be simplified by optimising for dedicated fiber
use.  Section \ref{lensletarray} then presents the design of the lens
array, comparing the macrolens design with other methods of close
packing fibers in the focal plane of the telescope.  By using
fore-optics to match the plate scale of the telescope with the lenslet
array, the spectrograph can then be used with other telescopes by
changing a single lens. 

The layout of SPIRAL Phase A is given in Section \ref{layout} along with
the optical design for the fiber-fed spectrograph. Section
\ref{fiberposition} briefly covers the performance and accuracy of the
fiber positioning technique that is used in the construction of the lens
array and focal plane unit for SPIRAL Phase A. The performance of the
spectrograph is given in Section \ref{char} and preliminary scientific
results are then presented in Section \ref{results}.

\section{AN IDEALISED FIBER-FED SPECTROGRAPH\label{ffs}}

The internal transmission of an optical fiber is predominantly dependent
on the wavelength of incident light and the focal ratio put onto the
fiber. The material in the core of the fiber contributes absorption
bands to the transmission properties beyond 700nm, predominantly from
hydroxyl ions (see \cite{Sch98}). With recent low water content
fibers, visible broad band transmission for lengths approaching many tens
of meters and efficiencies of over 90\% are available (\cite{Bar98}). 

Another dominant effect is focal ratio degradation (FRD), which has been
characterised and analysed in great detail (see \cite{Bar88},
\cite{Cla89}). Focal ratio degradation causes light entering a fiber at
a given focal ratio to emerge into a larger cone-opening angle. FRD
always acts to increase the etendue (A-$\omega$ product) of light due to
microbends and defects from an ideal cylindrical waveguide.  The ideal
focal ratio for most fibers is in the f/3 to f/6 range (\cite{Ram88}).

By taking FRD into account, efficiencies nearing those for an uncoated
fiber (92\%) have been measured by constructing the receiving optics to
accept a faster beam than is inputted into the fiber (\cite{Owe98} and
\cite{Ras93}). Properties of optical fibers are detailed in several
reviews (e.g. \cite{Bar88}; \cite{Bar98}).

\placefigure{fig5}

To describe the IFS fiber setup, a single fiber fed with
light from the telescope is shown in Figure \ref{fig5}. This can be
broken down into two sections - a focal plane unit (FPU) that feeds
light from the focal plane of the telescope into the fiber, and a
spectrograph that disperses light from the fiber and images the spectrum
onto the CCD camera.
The angle of sky subtended by an individual fiber, $\alpha_{fiber}$, is
determined by the focal ratio of the light entering the fiber,
$f_{fiber}$, the diameter of the fiber, $D_{fiber}$, and the diameter of
the telescope, $D_{tel}$. $\alpha_{fiber}$ can therefore be written as

\begin{equation}\label{eq1}
\alpha_{fiber} = \frac{D_{fiber}}{D_{tel}.f_{fiber}},
\end{equation}
where $D_{fiber}$ and $D_{tel}$ are in the same units and $\alpha_{fiber}$
is in radians.

From conservation of etendue in optical systems (sometimes known as the
A.$\Omega$ product) this result can be extended throughout successive
optics to give the angular extent of the fiber on the sky, regardless of
the number of optics in front of it. From Equation \ref{eq1}
$\alpha_{fiber}$ is independent of the size of the optic used to feed it
- it is only dependent on the focal ratio of the input beam. 

The Anglo-Australian Telscope has a diameter of 3.9 metres, so for a
fiber matched to 0.5 arcseconds on the sky and a core diameter of $50
\micron$, this requires an input focal ratio of $f/5.0$, which is a good
compromise between spectrograph cost (faster optics are more expensive)
and fiber efficiency (FRD is more detrimental to throughput at slower
focal ratios).

Optics in the FPU re-image the star or telescope pupil onto the end of
the fiber with a focal ratio of $f_{fiber}$. The input focal
ratio of the spectrograph (which is identical to the focal ratio of the
collimator, $f_{coll}$) should equal the focal ratio of the light
entering the fiber, i.e. $f_{coll}=f_{fiber}$, representing an ideal
fiber with no FRD. This is the case for slit-based spectrographs, where
the slit replaces an ideal fiber. For fiber-fed spectrographs
the collimator is made slightly faster to accomodate the intrinsic FRD
of the fiber ($f_{coll}<f_{fiber}$).

From the collimator of the spectrograph, the light then passes through
the dispersing element and camera with focal ratio $f_{camera}$ to form
a fiber spectrum.  The diameter of the image of the fiber on the
detector is:

\begin{equation}\label{eq2}
d_{image} = \frac{D_{fiber}.f_{camera}}{f_{coll}},
\end{equation}

where $D_{fiber}$ is the diameter of the core of the fiber in microns,
and $d_{image}$ is the diameter of the fiber image in microns.

The simplest FPU is one with no re-imaging optics and $f_{fiber}$ is
equal to the focal ratio of the telescope, $f_{tel}$.  This design is
used in many multi-object spectrographs where fibers have to be packed
closely together on the focal plane of the telescope in order to sample
crowded science fields, such as 2dF (\cite{Tay97}), Autofib-2
(\cite{Bri98}) and FLAIR (\cite{Par98}). This is not an ideal coupling
of the telescope to the optimal fiber focal ratio, and for other
applications fore-optics in the form of one or two lenses are introduced
between the focal plane of the telescope and the fiber ends to change
the focal ratio input put into the fibers.

Many of the first fiber-feeds were built as an add-on to existing
slit-based spectrographs, resulting efficiencies much lower than
expected.  By designing the spectrograph around the properties of the
optical fibers, a more efficient spectrograph could be built
that exploits their versatile image reformatting capabilities.

\section{LENSLET ARRAYS\label{lensletarray}}

Covering the focal plane of the telescope with a close packed array of
lenslets allows bi-dimensional spectroscopy of an extended region of the
sky. These lenslets tesselate to form a continuous pattern, with the
fibers reformatting the light from the lenslets into a fiber slit to be
fed into the spectrograph. 


In order to optimally sample the spatial structure of an astronomical
object, the plate scale of the lenses is such that the angular diameter
of the fiber is half the FWHM diameter of the telescope point spread
function.  Given that the seeing disk diameter on most telescopes is
typically 0.5 arcseconds and the plate scale of most large telescopes is
on the order of hundreds of microns per arcsecond, the expected lenslet
diameter is typically $350\micron$. This situation worsens in large
adaptive optic telescope systems where the lenslet size tends to smaller
image scales.  In the cases where a direct match to the plate scale of
the telescope is needed, a microlens array is used (see Figure
\ref{fig6}).

\placefigure{fig6}

Microlens arrays have a lenslet pitch of typically $150 - 500\micron$
and differ from macrolens arrays (see below) in that their lenslet
surfaces are manufactured in bulk on an optical substrate.  One method
involves the use of a metal template repeatedly stamped into a suitable
substrate, whilst another method uses the controlled curing of epoxy
resin on a suitable optical flat to form the lenslet surface. These give
rise to square or hexagonal close packed arrays that have a highly
regular spacing between lenslets, but the manufacturing process makes
them prone to large levels ($\sim 10\%$) of scattered light
(\cite{Lee98}). 

Alignment of the fibers with the microlens array is done by use of a
regular array of steel ferrules that match the lenslet pitch. Fibers are
fixed into the steel ferrule array and after optical polishing the
fibers in the ferrule array are registered with the lenslet array.  This
method has been adopted by many groups, especially where IFS is being
considered as an add-on to an existing spectrograph and the microlens
array needs to match the existing plate scale of the telescope, such as
the SMIRFS-IFS for the CGS4 spectrograph (\cite{Hay98}). 

From Equation \ref{eq1} the angular extent of the fiber on the sky is
dependent on its input focal ratio but independent of the size of
lenslets used in the FPU. Thus the lenslet array pitch is not
constrained and a macrolens array with a focal reducer can be used in
place of a microlens array. This technique was first suggested in the
early 1980's (\cite{Cou82}), pioneered in Russia on the SAO 6-meter
telescope (\cite{Afa90}) and is the one we use in SPIRAL Phase A.  

Macrolens arrays are lenslet arrays manufactured from individual glass
singlet or doublet lenses, with typical array pitches of 1mm - 5mm.
These lenslets are cemented onto an achromatic flat whose optical
thickness is equal to the focal length of the lenslets.  Each fiber is
then individually aligned with its respective lenslet, as opposed to
microlens arrays where optical alignment is dependent on the pitch
accuracy of the steel ferrule array holding the fibers. Macrolens arrays
have the potential for lower scattered light losses and by using
achromatic doublets greater wavelength coverage in the visible and near
infra-red can be gained. The relative merits of macro- and microlens
arrays are discussed further in other papers (\cite{Par97}).  

\section{OPTICAL LAYOUT OF SPIRAL PHASE A\label{layout}}

\placefigure{fig8}

Figure \ref{fig8} shows the layout of the spectrograph on the telescope.
A tilted aluminium mirror sits in the focal plane of the telescope. A
hole in the middle of the mirror allows light from the object to pass
through to the fore optics whilst the surrounding field of view is
imaged off the telescope axis by the target acquisition camera.  Light
passes through to to the fore optics and into the lens array. All of
these optical components are fixed onto an optical table in the
Cassegrain cage. Optical fibers connected to the back of the lens array
pass through the first strain relief box and out of the cage, down to
the spectrograph sitting on the floor of the observatory.  A second
strain relief box on the outside of the spectrograph leads to a fiber
slit where the fibers are reconfigured into a suitable straight slit.
The light is dispersed in the spectrograph and the fiber spectra imaged
on a CCD mounted inside a liquid nitrogen cooled vacuum dewar.

\subsection{Fore-optics and Lenslet Array}

\placefigure{fig7}

37 singlet lenses (PSK53A glass) spatially sample the sky in a close
packed hexagonal pattern. Each lenslet is 4\,mm across opposite corners
and the array measures 20\,mm in diameter. Optimum sampling
occurs when two spatial elements match the typical seeing of the
telescope (Nyquist sampling) and with a median seeing of one arcsecond.
On the AAT this corresponds to a scale of 0.5 arcseconds per lenslet,
giving a hexagonal array of 3.5 arcseconds across the corners.

Light from the telescope focus passes into a magnifying lens FOL1 which
produces an image of the telescope pupil (the primary mirror).  The
field lens in front of the array images all the fiber faces in the IFU
onto this common pupil (see Figure \ref{fig7}).  Each lenslet then takes
the section of sky image and projects an image of the pupil onto the end
of the fiber behind it.

The lenslet array is mounted on an achromatic flat composed of 4.46\,mm
of SF6 flint glass and 29.77\,mm of PSK53A Crown glass. The effective
focal ratio of the lenslets on the flat is $f/5.0$. 37 lenslets were
selected from a batch of 50, selected on the basis of surface quality
and focal length.

The image of the AAT telescope pupil completely fills the $50\micron$
core of the fiber. However, by adjusting the power of the magnifying lens
FOL1, the sampling on the sky can be changed. By making the image scale
less than 0.5 arcseconds per lenslet, the image of the telescope pupil
becomes smaller and under-fills the fibers. This allows smaller scales
to be observed without loss of efficiency. In SPIRAL Phase A the scale
is set to 0.5" per lenslet giving a total field of view of 3.5
arcseconds.

\placefigure{fig10}

The fibers in the fiber slit correspond to lenslets in the array in a
pre-defined pattern, so that images of the sky can be reconstructed from
reduced fiber spectra (see Figure \ref{fig10}). The fiber slit is
constructed so that fibers in the slit come from adjacent areas on the
sky. Any cross-talk introduced by the extraction of overlapping fiber
spectra are subsequently from adjacent areas in the observed field,
preventing spurious introduction of artefacts in the reconstructed
integral field image. 

In extreme cases of cross-talk this results in an elongation of extended
objects in the reconstruction.  Since the SPIRAL spectrograph is
designed to be used with 500 fibers in the Phase B fiber feed, the 37
fibers in the prototype were widely spaced apart along the length of the
imaging area of the spectrograph slit and there was no cross-talk.

\subsection{The Fiber Feed}

The SPIRAL fibers were from Polymicro and made from fused silica. The
core diameter was $50\micron$, with a lower-index doped fused silica
cladding out to $70\micron$ diameter and a polyamide coating up to $90\micron$. A further environmentally protective layer of soft
polyethylene prevented damage of the fiber during handling, but was
easily removed by soaking in acetone prior to fixture in the instrument.

A slow curing two-component epoxy mechanically holds the fiber within a
thin stainless steel tube. Successively larger steel tubes were added
and fixed in place to form a composite stainless steel tube with an
outer diameter of $500\micron$ and a single fiber held rigidly in
the center.

The 37 fibers then pass into a protective polyamide tubing which guides
the fibers with minimum bending into a tough, steel wound plastic
covered conduit.  This conduit passes through the Cassegrain cage of the
telescope and down to the floor of the observatory. One meter from each
end of the conduit is a strain relief box, where the fiber passes in an
unsupported loop across the inside of the box before passing through out
the other side. These two spare length boxes ensure that the expansion
and contraction of the flexible conduit does not stretch and break the
inelastic and delicate optical fibers.

The fiber passes into the fiber slit assembly. Here the 37 fibers are
rearranged into a straight line, the fibers spaced evenly along the slit
and parallel to each other. The fibers are held against a flat reference
plate by a metal block with accurately machined grooves in its surface.
The mechanical pressing action of the grooves and reference plate serve
as a vice with which to hold the fibers in place. All fibers lie
parallel to each other in a plane perpendicular to the surface of the
final fiber slit. Epoxy resin is used to fill the inter-fiber spacing
and provide a strong mechanical support for the fibers in the slit
block. Finally a thin cover slit was fixed with optical cement across
all the ends of the fibers in the polished fiber slit. In this way, any
scratches and imperfections in the ends of the fibers were filled in
with index-matching cement and the exposed surface of the cover slip was
coated with an anti-reflective layer of magnesium flouride.

Our fiber feed design has no exposed optical fiber surfaces, an
important consideration when the active optical surface is $50\micron$
in diameter. Left unprotected, the smallest scratch or contamination
with dust can lead to a significant loss of efficiency. Both ends of the
fiber feed are mated against a glass surface with a liquid optical
cement, making an index matching interface and reducing scattering and
reflection losses. The addition of a magnesium flouride anti-reflection
coating to both ends of the fiber feed optics further improves the
throughput.

\subsection{The Spectrograph\label{spectrograph}}

The input focal ratio for the spectrograph is $f/4.8$, slightly faster
than the $f/5.0$ beam entering the fibers. This allowed for FRD
broadening of the beam. The spectrograph also had to produce
monochromatic images over the whole area of the Tek CCD, requiring
imaging of a slit length of 25\,mm.

The SPIRAL spectrograph was designed by Damien Jones to use the Tek
$1024\times1024$ pixel optical CCD. The $24\micron$ pixels of this CCD
give Nyquist sampling of the $50\micron$ fiber diameter, which defines
the entrance slit width to the spectrograph. Because this fixed slit
width is smaller than most slit based spectrographs and is independent
of the slit-adjusted seeing, the camera and collimator are combined to
make a double pass system with a total magnification of unity. This then
allows the grating to be used in a near-Littrow configuration with the
grating used at its highest theoretical efficiency. For a double pass
system the Littrow condition is exact only for object and image
superimposed on each other, but by separating the fiber slit and
detector beams by 22.5\,mm, the deviation from the Littrow configuration
is minimised. 

\placefigure{fig11}

The resultant design is shown in Figure \ref{fig11}.  The most suitable
optical configuration for the spectrograph is a field-flattened Petzval
system, otherwise known as a P1 Petzval configuration. This consists of
two separated positive doublets (L2 and L3) followed by a field
flattening singlet (L1) near the focal plane. A small reflex prism
brings the fiber slit image close to the optical axis of the system, and
the grating G1 sits in the collimated space beyond L3. 

The design also allows for rotation and tilt of the dewar, eliminating
the need for an extra field correcting lens and thus reducing the cost
of the system - a plot of chromatic focal shift (Figure \ref{fig12})
shows the extent of refocusing needed across the wavelength range of the
spectrograph. 

\placefigure{fig12}

\section{FIBER POSITIONING\label{fiberposition}}

\subsection{Mechanical Alignment of Fibers with Individual Lenslets}

The macrolens array allowed individual positioning of fibers on the back
of the lenslet array. Individual $50\micron$ diameter fibers were fixed
with epoxy resin into metal ferrules 10\,mm long and with an outer
diameter of $230\micron$. Further steel ferrules were added to the outer
diameter of this ferrule to produce a composite ferrule fiber holder
2\,mm in diameter with the optical fiber held firmly along the ferrule
axis. All the fibers for the lens array were prepared this way and then
placed together in a polishing jig. This jig was then used to
simultaneously polish all 37 fiber faces and ferrules to optical
quality.

The field lens FOL2 from the fore-optics was fixed in front of the
lenslet array and a graticule was placed in the focal plane of the field
lens with the cross-hairs centered on the optical axis of the field lens
(see Figure \ref{fibposition}). An electronic camera with a narrow-band
filter and a microscope objective was focused on the cross-hair
graticule. In a perfectly constructed IFU, the field lens brings all the
individual fiber images from the lenslet array and brings them to a
common focus on the center of the graticule. By back illuminating each
fiber and using an x,y,z stage to manoeuvre the fiber ferrule, each fiber
could be individually aligned with the optical axis of its respective
lenslet.

\placefigure{fibposition}

When the fiber image was centered in the gratiule, the fiber was
correctly aligned. The optical cement was applied to the face of the
fiber to fix it to the back of the lens array, and any dust or air
bubbles that existed between the fiber and lenslet array could be
clearly seen in an electronic camera image and removed. When the fiber
was ready, a UV source was used to cure the cement and fix the fiber in
place.

After all 37 fibers were positioned, the lenslet array was assembled
into an aluminium protective housing and a slow curing potting compound
was poured onto the back of the lens array and in between the fiber
ferrules. This increased the mechanical support of the ferrules and
prevented sudden mechanical shocks from breaking the UV cement bond
between ferrule face and lenslet array.

\subsection{Measured Positioning Errors\label{fpe}}

In the fore-optics, lens FOL1 forms a pupil image of the telescope. This
pupil is then re-imaged with FOL2, through each individual lenslet and
forms a pupil image on the fiber face. If light is shone back up the
fibers to illuminate all the lenslets, an image of each
fiber is formed at the position of the telescope pupil - if all the
fibers are concentric with the optical axis of their respective lenslet
and all 37 fiber face images overlap exactly at the focus of FOL2. 

If the fibers are not perfectly aligned then the fiber images do not
overlap exactly, and each fiber image has a measured displacement from
the mean position of all the combined images. From the image scale
of the fiber image, the positional error in the fiber plane can be
calculated.

We define the decentering error to be the radial distance of one fiber
from the mean fiber position of the whole fiber feed. Assuming the
positioning errors to be random and independent of each other, all the
decentering errors form a Gaussian distribution around a mean fiber
position and we can define an RMS decentering error for all the fibers
that gives a measure of all the fiber misalignemnts in the lenslet
array.

After the SPIRAL IFU was assembled, a measurement of the fiber positions
was made using the alignment apparatus and the graticule cross hairs.
The RMS decentering error was $10\micron$ with a maximum error of
$13\micron$ (see Figure \ref{fiboffs}). This value was considerably
larger than expected, and disagreed strongly with the value estimated by
eye during the assembly of the lenslet array. 

The reason for this large error was that the potting compound had
displaced the fiber ferrules and caused a misalignment of the ferrules
as it cured.  Further investigation revealed that the optical cement
used to hold the fibers in place on the lenslet array required
considerably longer exposure to the UV source than had been suggested in
earlier experiments This resulted in the fibers not being held firmly on
the back of the lens array, and subsequently led to a calculated
reduction of transmission efficiency of $20\%$ in the fiber feed due to
misalignment of the fibers and lenslets.

\placefigure{fiboffs}

The lenslet array could not be rebuilt without causing considerable
damage to the fibers, so we used the prototype in its current state. The
alignment technique was more accurate than the SPIRAL lenslet array, and
the same method was used in the construction of a 100 element infra-red
IFU for the Cambridge OH Suppression Instrument (COHSI) (\cite{Enn98}). 

The technique was vindicated when the constructed lenslet array of 100
lenslets was measured to have an RMS decentering error of $3\micron$
(\cite{Ken98b}).

\section{PERFORMANCE OF THE FIBER FEED AND SPECTROGRAPH\label{char}}


\placefigure{fig14}

Figure \ref{fig14} shows a data frame taken with the SPIRAL Phase A
spectrograph. An overscan and zero level frame has been subtracted, and
cropped to remove stray and reflected light from the edge of the image.
This exposure clearly shows absorption bands in the twilight sky.


\subsection{Point Spread Function}

The point spread function across the CCD was measured by fitting
Gaussian profiles to an arc lamp frame containing many narrow emission
lines. These proved ideal for measuring the PSF introduced by the
spectrograph optics. 

The PSF was measured to be $2.31\pm0.11$ pixels, corresponding to an
image size of $69.3\pm0.3\micron$, larger than the fiber core diameter
of $50\micron$, but in agreement with the PSF of the optical model as
produced with commercial ray tracing software.  There was no significant
variation in the measured PSF across the image.

\subsection{Spectral Range and Resolution}

Different gratings were inserted into the spectrograph and arc spectra
images were taken for various wavelength ranges. By using a dome flat
field to define the fiber tracks the spectra were extracted and a
dispersion solution identified for the arc lines. The best low-order fit
for the dispersion was a 4th order polynomial. The results for three
configurations are shown in Table \ref{table1}.

\placetable{table1}

\subsection{Total Transmission}

Observations of spectrophotometric standard stars provided an estimate
of the throughput of the spectrograph and fiber feed.  A two-dimensional
low-order polynomial was fitted to a standard star data frame to remove
scattered light.  The frame was then summed along the dispersion axis to
form a one-dimensional spectrum which was dispersion corrected and
wavelength calibrated.

Over three nights the SPIRAL spectrograph was calculated to have an
efficiency of between 10-15\% with an estimated median value of 12\% at
6250\AA~.  The main variation in throughput was due to the fiber
misalignment discussed in Section \ref{fpe}. By estimating and measuring
the throughput of all the optical components in the system, the
throughput of the IFU and fiber feed throughput was estimated to be
$57\pm3\%$ (see Table \ref{table3}). The throughput of the fore-optics,
IFU and fiber feed combined is $52\pm3\%$, but if there was no fiber
misalignment then this number increases to $65\pm3\%$.



\placetable{table3}

\section{FIRST RESULTS FROM SPIRAL A \label{results}}

At the beginning of our observing the MIT/Lincoln Laboratory
$2048\times4096$ pixel CCD became available for use in the spectrograph.
With its higher quantum efficiency and lower readout noise, we used this
for the observations. Since this CCD has 15$\micron$ square pixels the
MIT/LL was set to 2 by 2 pixel binning, forming a effective frame of
1024$\times$2048 $30\micron$ pixels.

\subsection{Supernova 1987A}

To test the integral field mode of SPIRAL, an object with small scale
detail was required for our 3.5 arcsecond field. The Large Magellanic
Cloud SN 1987A was observed during sub-arcsecond seeing on 1997 November
23.

\placefigure{fig16}

The circumstellar environment of the supernova is still not well
understood (\cite{Cro97}). An HST optical image with the SPIRAL field of
view is shown in Figure \ref{fig16}. The three ring structure which is
visible in the image lies on the surface of an hourglass shape inclined
partially to our line of sight (\cite{Cro89}). The inner ring marks the
waist of the hourglass (\cite{Cro91}) whilst the two larger rings mark a
circumference further up the lobes, delineating the extent of a
double-lobe nebula (\cite{Cro95}).

The observations consisted of two sets of data frames. Three 1800 second
exposures were taken of the supernova along with three 1200 second
exposures of the adjacent sky. This was especially important
as the supernova is within an HII region, with the associated nebular
emission from various lines such as [\ion{O}{3}] and [\ion{N}{2}].

Figure \ref{fig17} shows [\ion{O}{1}]$\lambda6300$,
[\ion{N}{2}]$\lambda6548$, H$\alpha\lambda6563$,
[\ion{N}{2}]$\lambda6583$, [\ion{S}{2}]$\lambda6716$ and
[\ion{S}{2}]$\lambda6731$ emission lines. The supernova spectra have
been background subtracted. However, some sky emission lines still
remain from incorrect background subtraction.

\placefigure{fig17}

Nebular emission lines in the LMC are red shifted by its recession
velocity (typically 300\,km\,s$^{-1}$) and are visible in the spectra.
The bright narrow H$\alpha$ line is predominantly from the supernova and
this line is mapped in Figure \ref{fig18}. A lower intensity profile
broadening of $\sim$2600\,km\,s$^{-1}$  is visible in the spectrum. This
is identified as the first signs of the fast moving supernova ejecta
hitting and shocking slow moving circumstellar gas released during the
progenitors giant star phase (\cite{Che92} and \cite{Son97}). 

Four maps have been generated which show the structure in the region of
the supernova (Figure \ref{fig18}). The bulk of the continuum emission
is from Star 2, in the South-east of the field of view. Selecting a
similiar bandpass to the HST filter shows similiar intensity structure,
and by fitting profiles to the supernova wind, the location of the
supernova is confirmed. An [\ion{N}{2}] map shows an inner ring that has
been shock-ionised by the initial UV flash from the explosion.

\placefigure{fig18}

\subsection{Detection of a Brown Dwarf}

To test the high resolution capability of the spectrograph several brown
dwarf candidates were examined. One of the simplest tests for young
brown dwarfs is detection of the Lithium absorption line
\ion{Li}{1}$\lambda6707.8$ in very red objects with low absolute
magnitude. This implies that the core temperature of the star has not
reached the Li/H burning temperature and therefore not reached a higher
H/H burning temperature (\cite{Reb92}). A more detailed discussion about
the selection procedure can be found in (\cite{Ken98}).

\placefigure{figbd}

A list of candidate stars was provided by Hugh Jones (private
communication) and the first star observed was LP 944-20. The
observation was 1800 seconds with the 1200R grating used in second
order, providing a calculated spectral resolution of 15000.  As of
mid-1997, when these observations were undertaken, there were only a
small number of known brown dwarfs. LP 944-20 showed strong \ion{Li}{1}
absorption (see Figure \ref{figbd}) thus confirming its membership to
this select group. However, during our observations a spectra from the
CASPEC echelle spectrograph and ESO telescope was published
(\cite{Tin98}) showing \ion{Li}{1} absorption.

This now provided an excellent opportunity to compare the accuracy of
the SPIRAL spectrograph with a standard slit-based spectrograph (see
Table \ref{table5}).

\placetable{table5}

Both methods to determine the apparent magnitude and Lithium absorption
agree within errors, with the only disagreement being the heliocentric
velocity. Tinney derived the velocity by cross-correlation with a
standard star, whilst the SPIRAL estimate was made with the centroid of
the Lithium line only, resulting in the larger error estimate of the
SPIRAL spectrograph.

\section{SUMMARY\label{conclusions}}

We have manufactured a lenslet array from individual lenslets and
demonstrated that fibers can be aligned with these lenslets to
$3\micron$ RMS error, providing an alternative technique to the
microlens array method. The use of fore-optics makes the fiber feed
portable and the use of fibers leads to a simplified and more efficient
spectrograph design.

Integral field observations of a supernova remnant allow maps of spatial
variations in spectral line properties to be generated, and high
spectral resolutions to be achieved. The success of this prototype led
to the construction of SPIRAL Phase B and its comissioning in May 2000.

\acknowledgments 

We are indebted to the Staff and Technicians of the Anglo-Australian
Observatory, who were always ready to help out in getting the instrument
on the telescope. MAK thanks Karl Glazebrook and Dave Lee,
who provided invaluable help with the software and calibration on the
observing runs, and to Richard Ogley for proof-reading this paper.

\clearpage

\clearpage
\begin{deluxetable}{ccccc}
\footnotesize
\tablecaption{Measured dispersions and resolutions of the SPIRAL spectrograph\label{table1}}
\tablewidth{0pt}
\tablehead{
\colhead{Grating} & \colhead{Wavelength Range} &
\colhead{Dispersion} & \colhead{Line FWHM} &
\colhead{Spectral Resolution} \\
\colhead{} & \colhead{(\AA)} &
\colhead{(\AA/pixel)} & \colhead{(\AA)} &
\colhead{(R)}
}
\startdata
270R 1st order &2400&1.525&$2.820\pm0.180$& 2730 \\
1200R 1st order &500&0.314&$0.864\pm0.036$& 7550 \\
1200R 2nd order &250&0.155&$0.419\pm0.017$& 15000 \\
\enddata
\end{deluxetable}

\clearpage

\begin{deluxetable}{lc}
\footnotesize
\tablecaption{Measured and estimated efficiencies for SPIRAL Phase A on
the AAT\label{table3}}
\tablewidth{0pt}
\tablehead{
\colhead{Optical Component} & \colhead{Transmission} \\
\colhead{} & \colhead{(\% at 6000\AA)}
}
\startdata
Atmosphere and telescope &$56\pm1$\tablenotemark{a}\\
Fore-optics &$92\pm1$\tablenotemark{b}\\
IFU and fiber feed &$57\pm3$\\
Spectrograph and 1200R grating &$53\pm1$\tablenotemark{c}\\
CCD&$77\pm1$\tablenotemark{c}\\
Total measured efficiency &$12\pm2$\\
\enddata
\tablenotetext{a}{Karl Glazebrook (priv. commun.)}
\tablenotetext{b}{Estimated for 4 air-glass surfaces}
\tablenotetext{c}{Dave Lee (priv. commun.)}
\end{deluxetable}

\clearpage
\begin{deluxetable}{lccc}
\footnotesize
\tablecaption{Comparison of measured properties of LP 944-20\label{table5}}
\tablewidth{0pt}
\tablehead{
\colhead{} & \colhead{Apparent Magnitude} &
\colhead{EW of \ion{Li}{1}} & \colhead{Heliocentric Velocity}\\
\colhead{} &
\colhead{($m_{R}$)} &
\colhead{(\AA)} &
\colhead{(km s$^{-1}$)}
}
\startdata
Value from literature&$17.1\phn\pm0.1$\tablenotemark{a}\phn&$0.53\pm0.05$&$+13\pm4\tablenotemark{b}\phn$ \\
Value from SPIRAL A&$17.19\pm0.10$&$0.54\pm0.06$&$-22\pm12$\\ 
\enddata
\tablenotetext{a}{\cite{Kir97}}
\tablenotetext{b}{Cross-correlation with reference star}
\end{deluxetable}

\clearpage 

\plotone{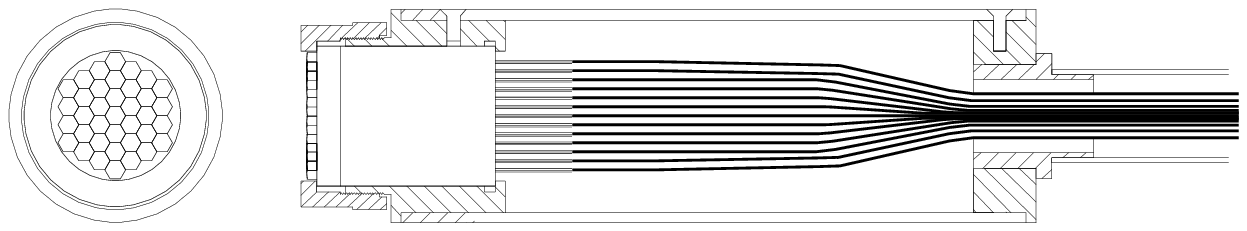} 

\figcaption[nfg1.eps]{The SPIRAL Phase A Integral Field Unit. The front
view (left) shows the arrangement of the 37 hexagonal lenslets and the
cutaway side view (right) shows the metal casing and fiber optics from
the lenslets leading down to the spectrograph.\label{fig1}} 

\clearpage

\plotone{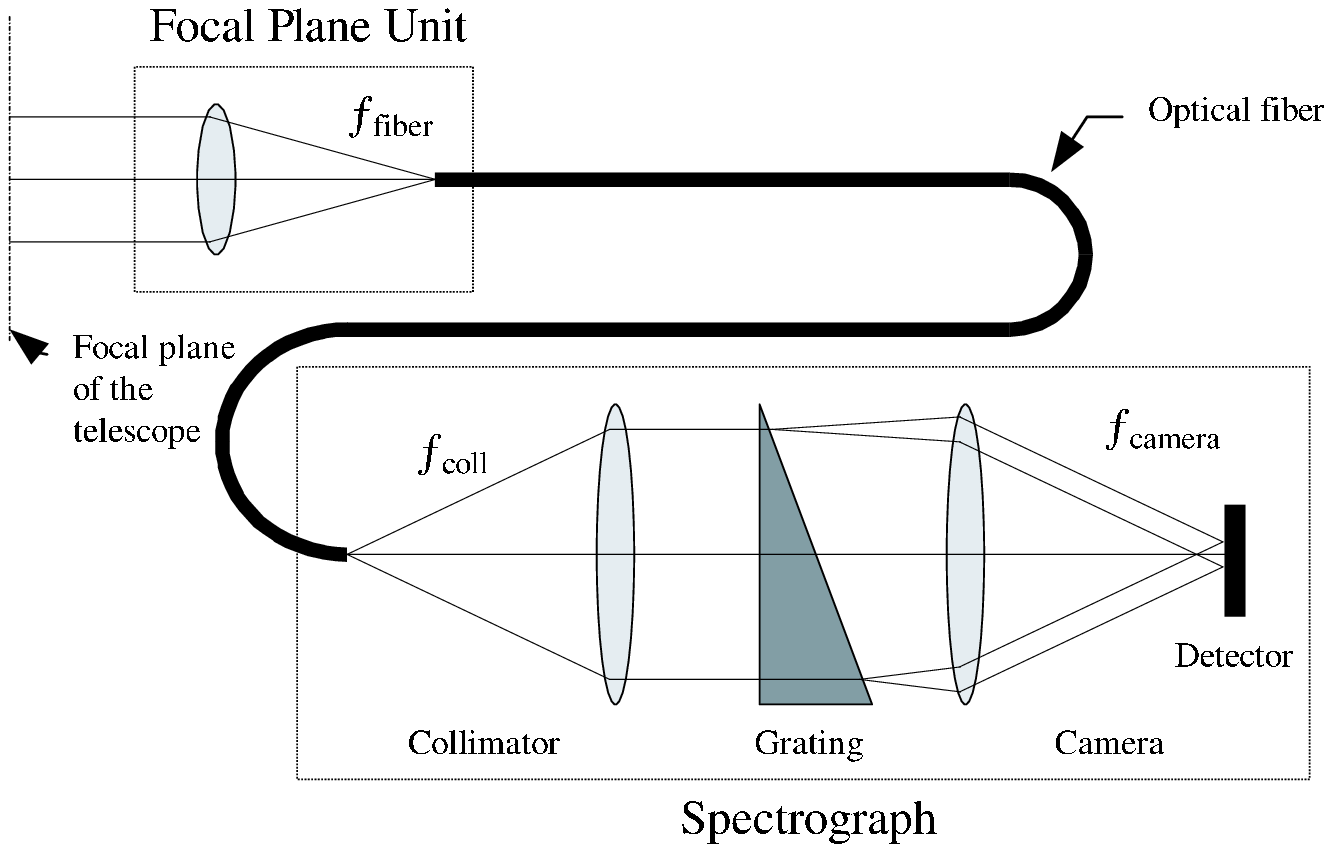}

\figcaption[nfg2.eps]{A spectrograph fed by an optical fiber. In this
case, the focal plane unit images the telescope pupil onto the core of
the fiber.\label{fig5}}

\clearpage
\plotone{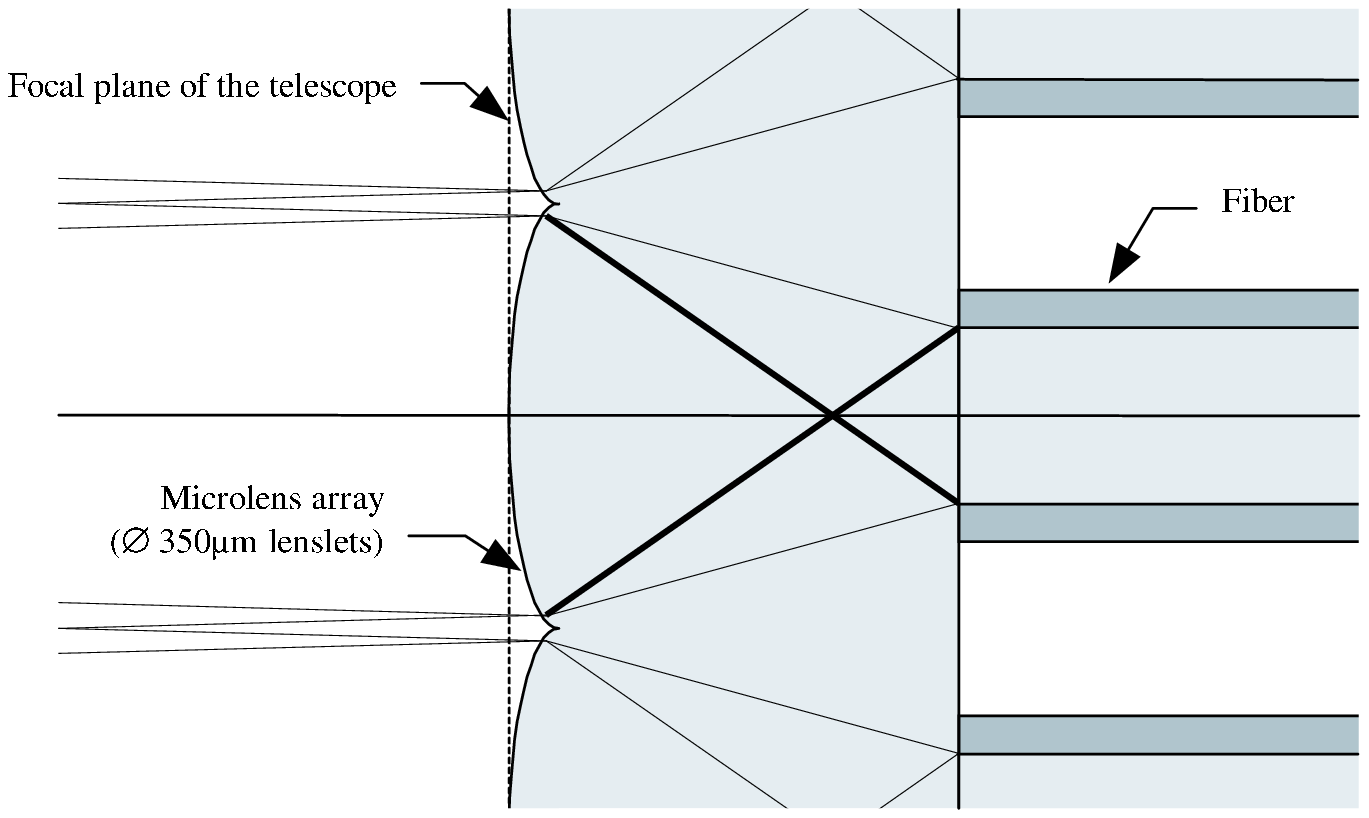}

\figcaption[nfg3.eps]{A microlens array feeding optical fibers. The
thicker rays from the middle lenslet show how the edge of the fiber is
fed with faster rays than those in the middle of the fiber.\label{fig6}}

\clearpage
\plotone{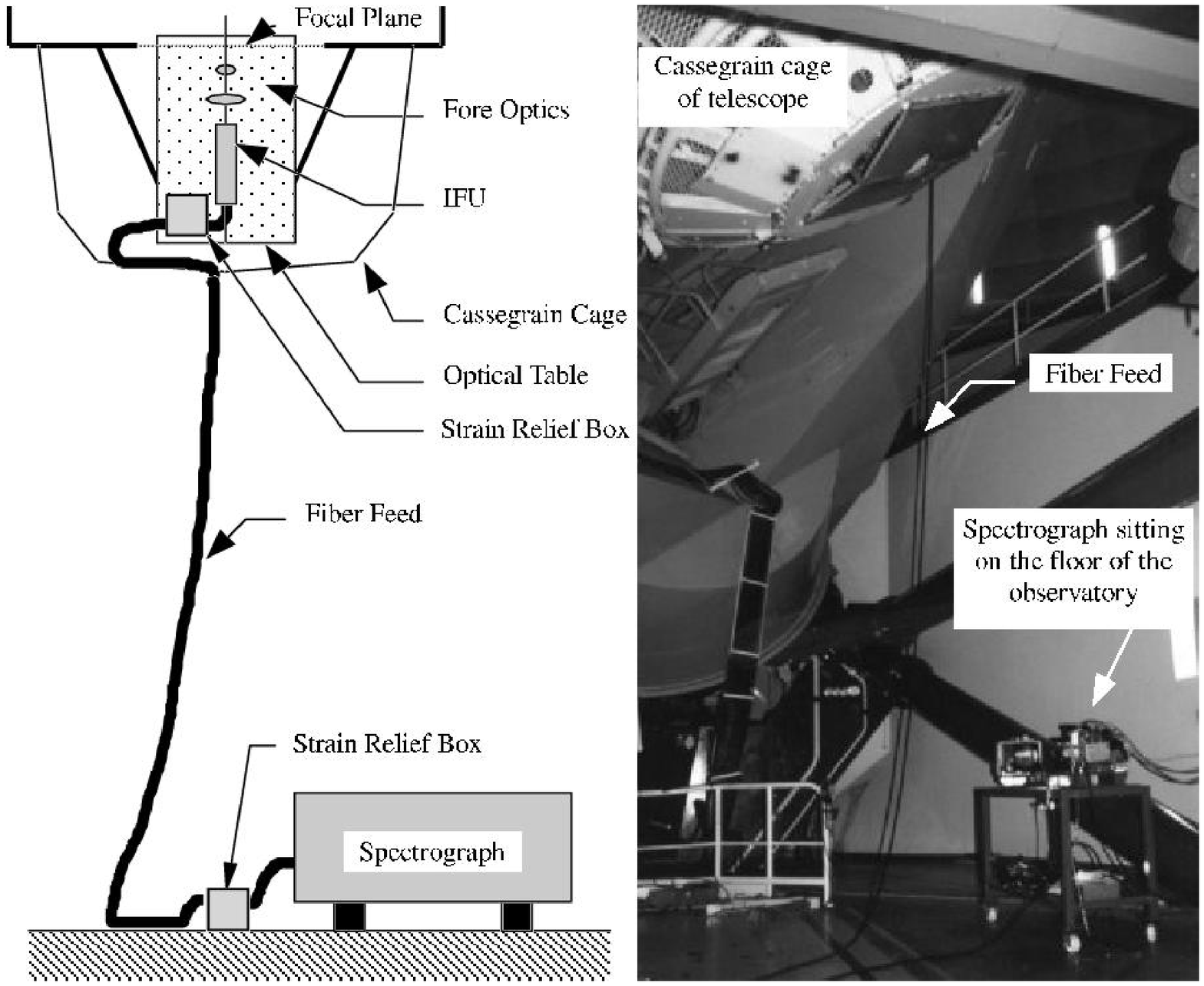}

\figcaption[nfg4.eps]{Layout of SPIRAL Phase A on the Anglo-Australian
Telescope. The spectrograph is shown with its light-proof covers
removed. The strain relief box for the fiber slit end is mounted
underneath the spectrograph framework.  \label{fig8}}

\clearpage
\plotone{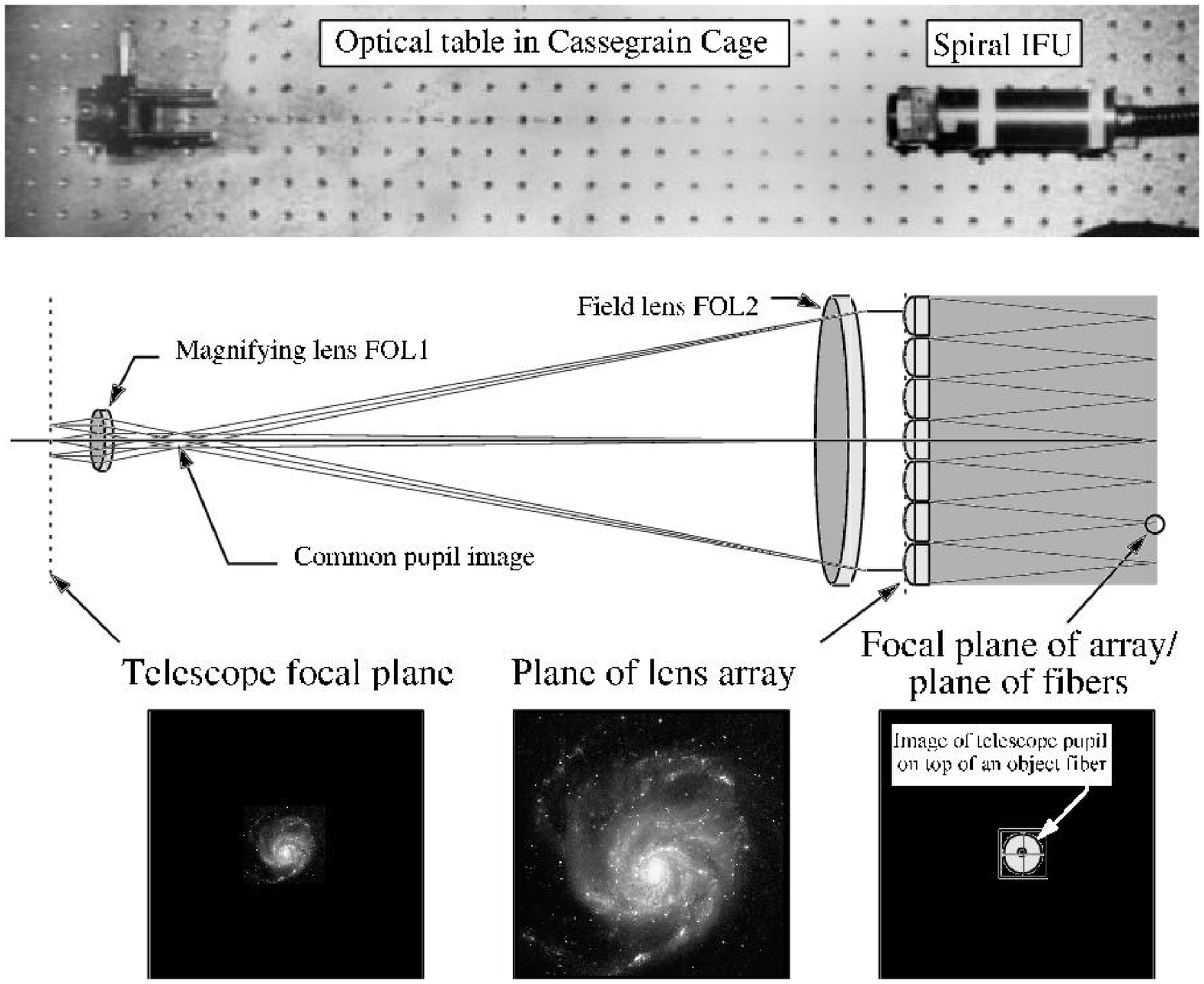}

\figcaption[nfg5.eps]{The fore-optics for the integral field mode. These
two lenses act to magnify an extended object (such as a galaxy) to match
the plate scale of the lens array with the seeing at the telescope. At
the focal plane of the fibers each lenslet projects an image of the
pupil onto the end of the object fiber.\label{fig7}}

\clearpage
\plotone{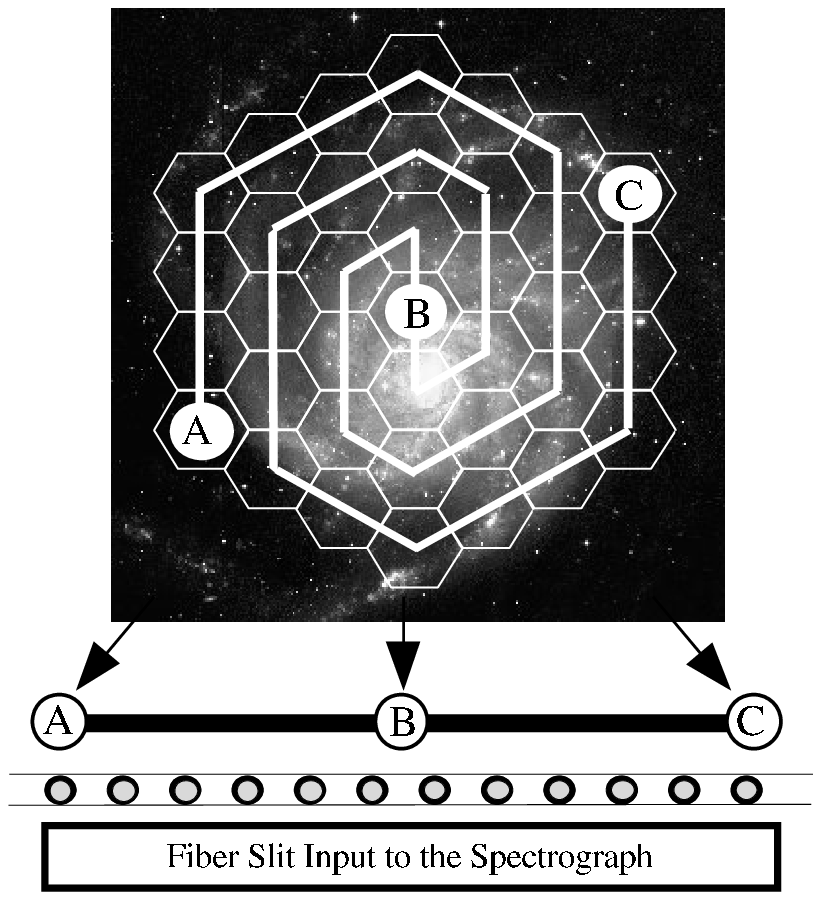}

\figcaption[nfg6.eps]{Schematic layout of the SPIRAL fiber slit. The
fibers are reformatted from a spiral pattern on the sky into a long slit
suitable for dispersion in a spectrograph. The spiral pattern ensures
that fibers adjacent on the slit are adjacent on the sky.\label{fig10}}

\clearpage
\plotone{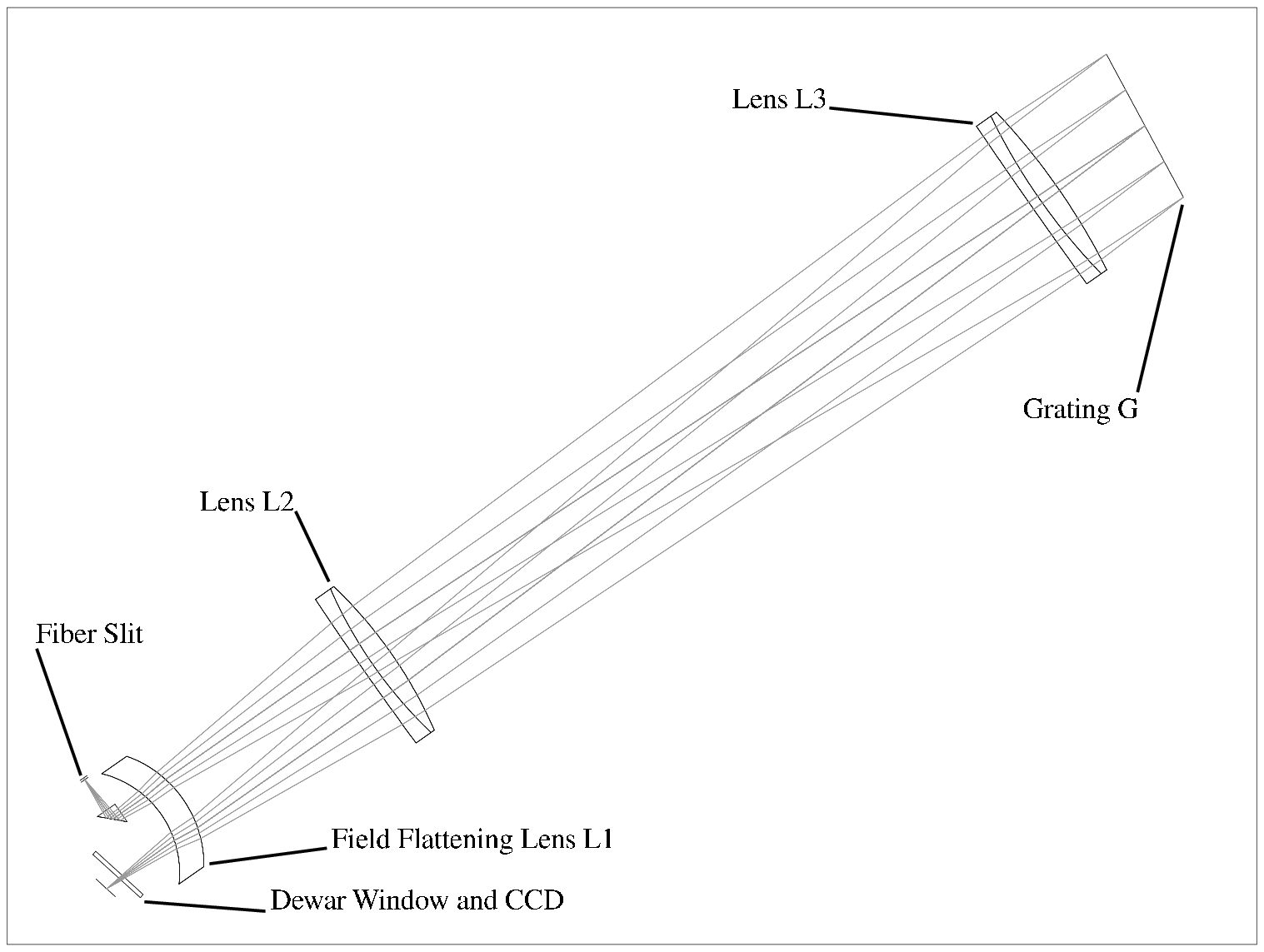}

\figcaption[nfg7.eps]{Optical layout of the SPIRAL spectrograph.
Monochromatic light emerges from the fiber slit in this ray-trace and is
folded by a prism into the Petzval system (L1,2,3). After being
diffracted by the grating G1 the light passes back through the system
and is focused on the detector inside the dewar (not
shown).\label{fig11}}

\clearpage
\plotone{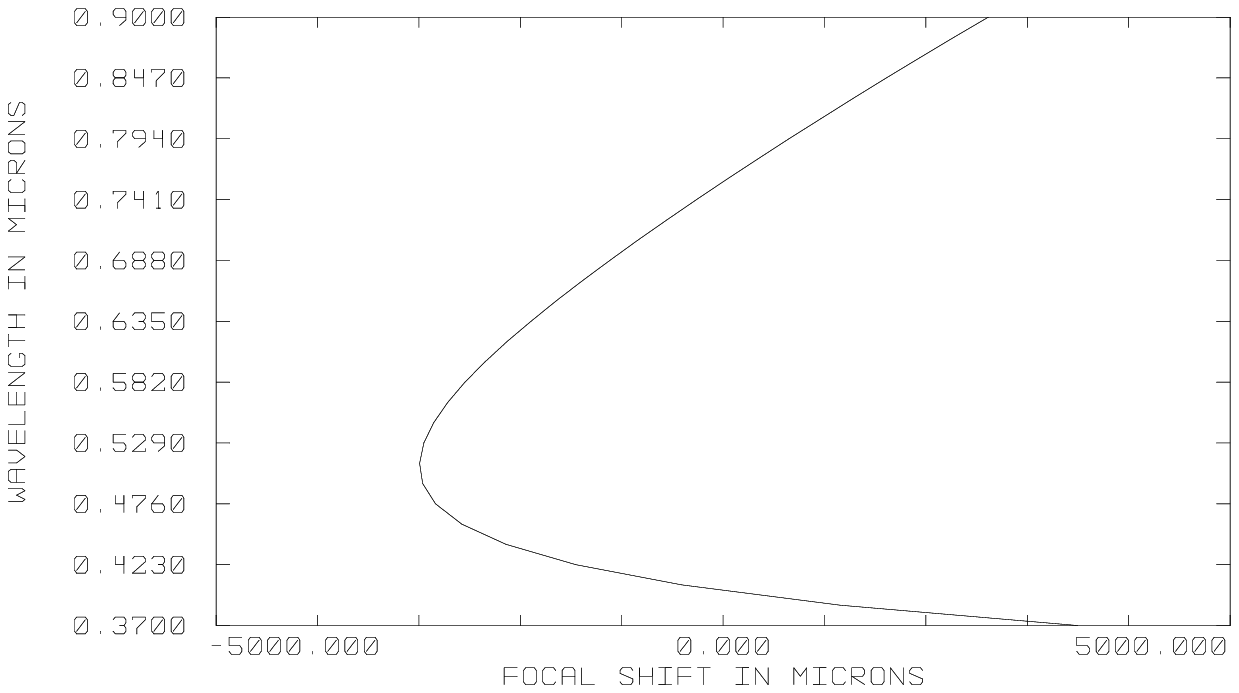}

\figcaption[nfg8.eps]{Chromatic focal shift for the SPIRAL
spectrograph. \label{fig12}}

\clearpage
\plotone{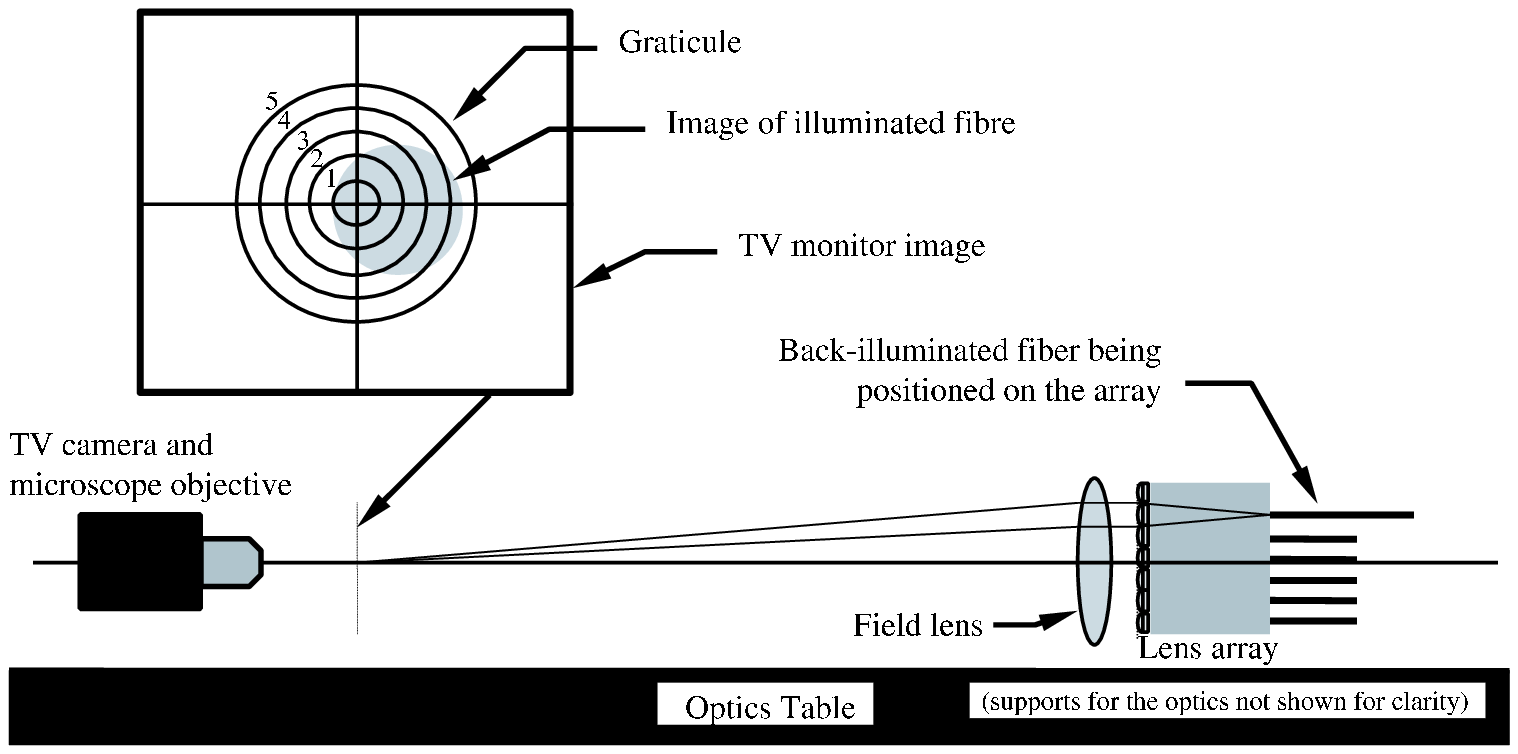}

\figcaption[nfg9.eps]{Optics used in the alignment of fiber
ferrules on the back of the lenslet array.\label{fibposition}}

\clearpage
\plotone{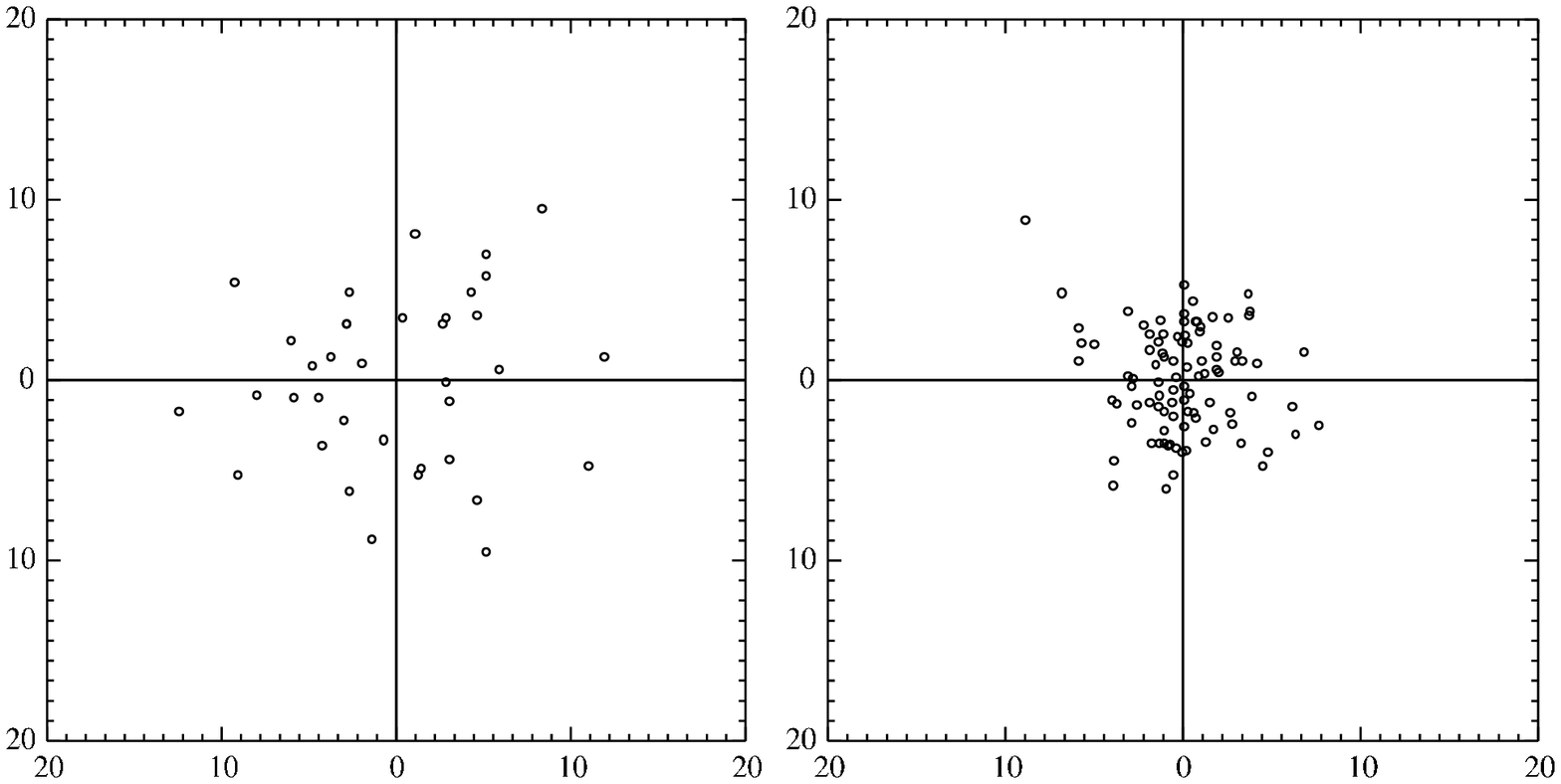}

\figcaption[nfg10.eps]{SPIRAL Phase A (left) and COHSI (right)
decentering errors. Both plots have scales measured in microns centered
on the mean fiber position.\label{fiboffs}}

\clearpage
\plotone{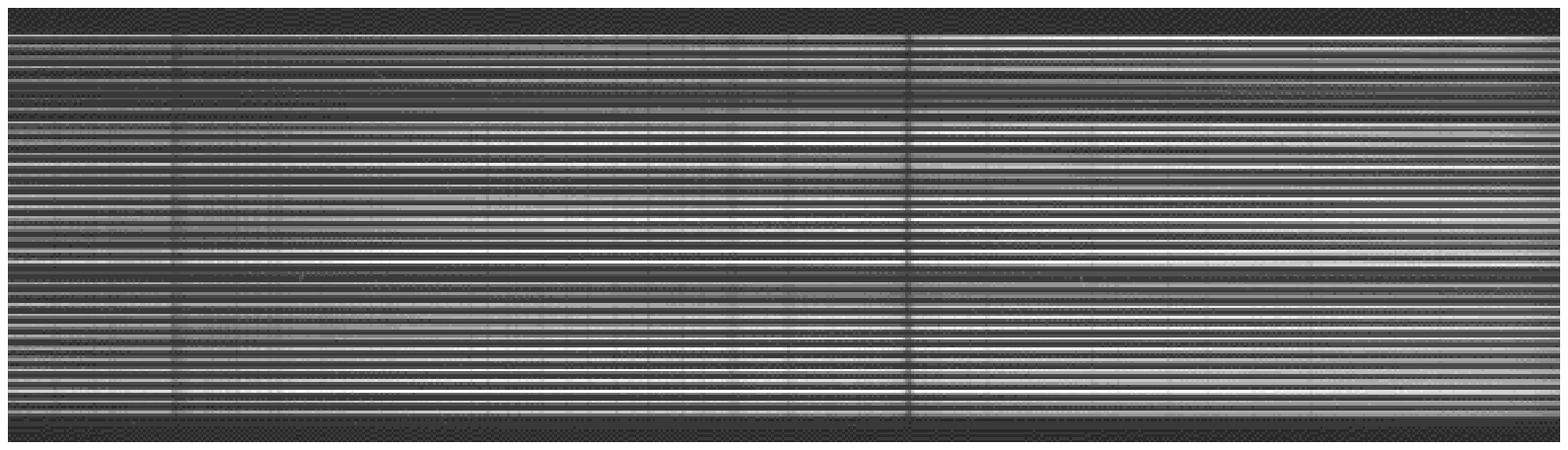}

\figcaption[nfg11.eps]{A raw IFS data frame. In this frame the
dispersion axis is across the page and the 37 separate fiber tracks can
be seen. This is a twilight sky exposure, clearly showing absorption
features in the atmosphere and the variation in throughput between
fibers.\label{fig14}}

\clearpage
\plotone{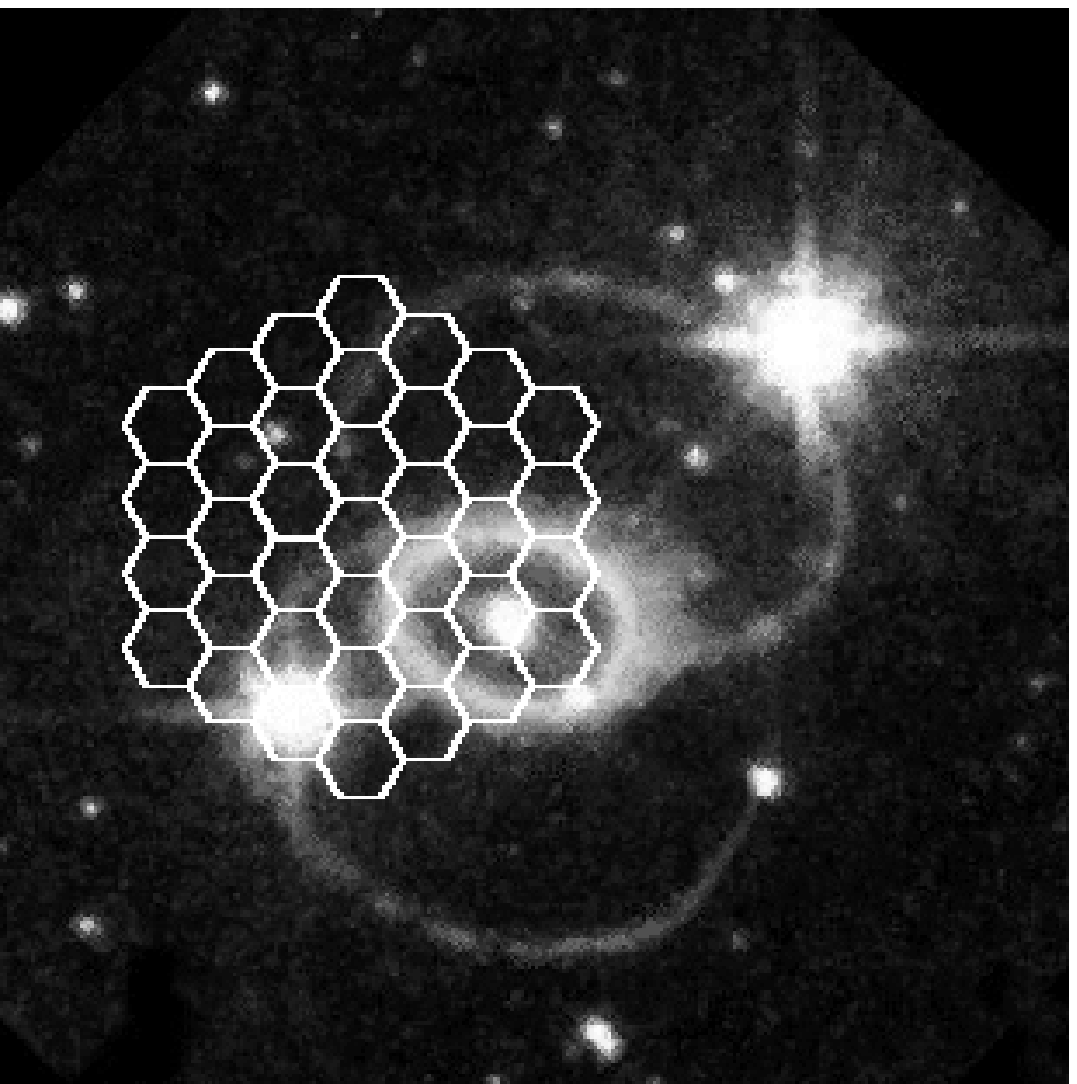}

\figcaption[nfg12.eps]{HST WFPC image of 1987A. The estimated position
of the SPIRAL lens array is shown superimposed on the image. North is to
the top and East to the left. (Photo: Chun Shing Jason Pun (NASA/GSFC),
Robert P. Kirshner, taken on March 5 1995 in \ion{N}{2}$\lambda6584$).\label{fig16}}

\clearpage
\plotone{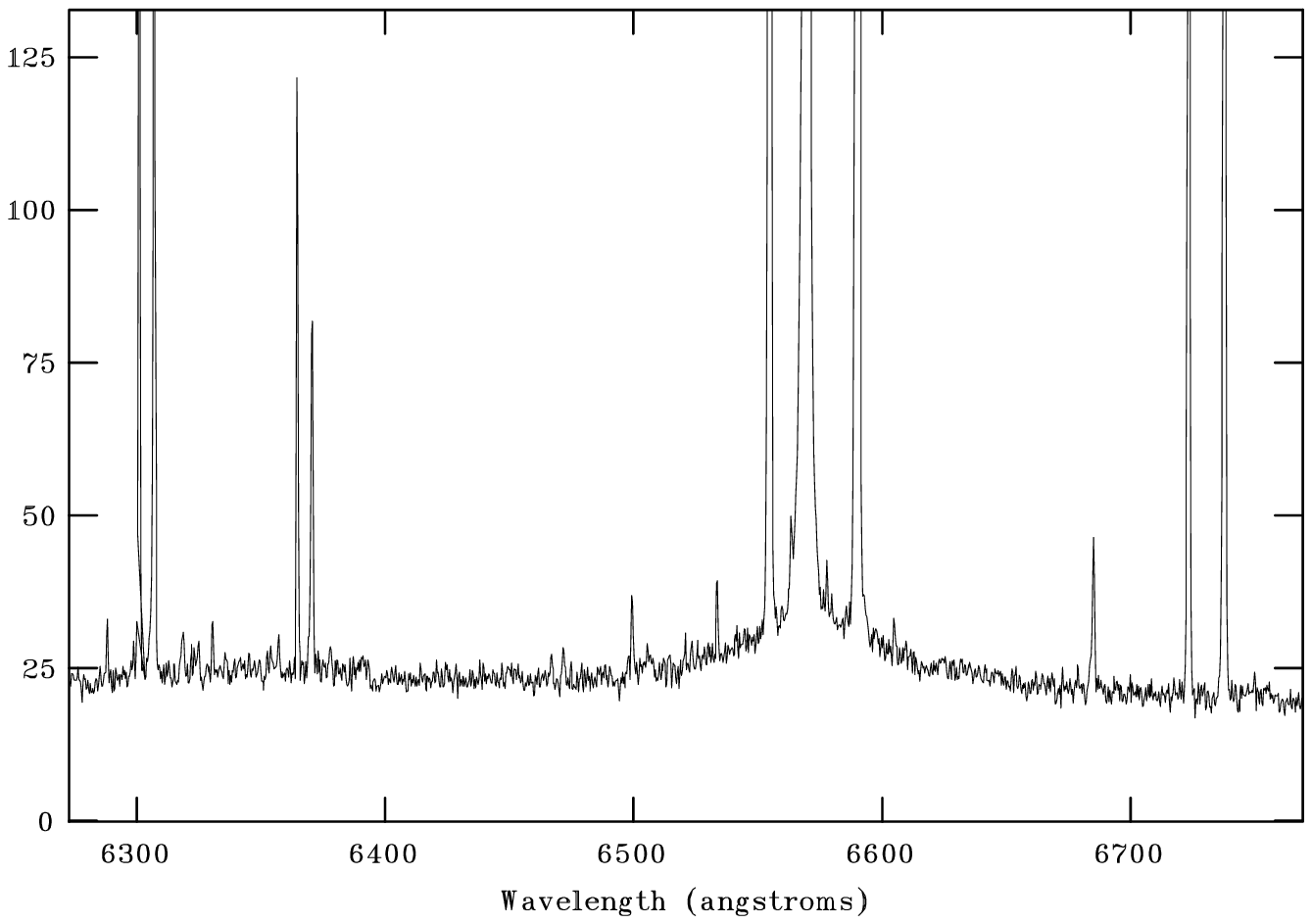}

\figcaption[nfg13.eps]{The SN 1987A spectrum. Superimposed on the bright
narrow hydrogen line is a broad line contribution from fast moving
ejecta.\label{fig17}}

\clearpage
\plotone{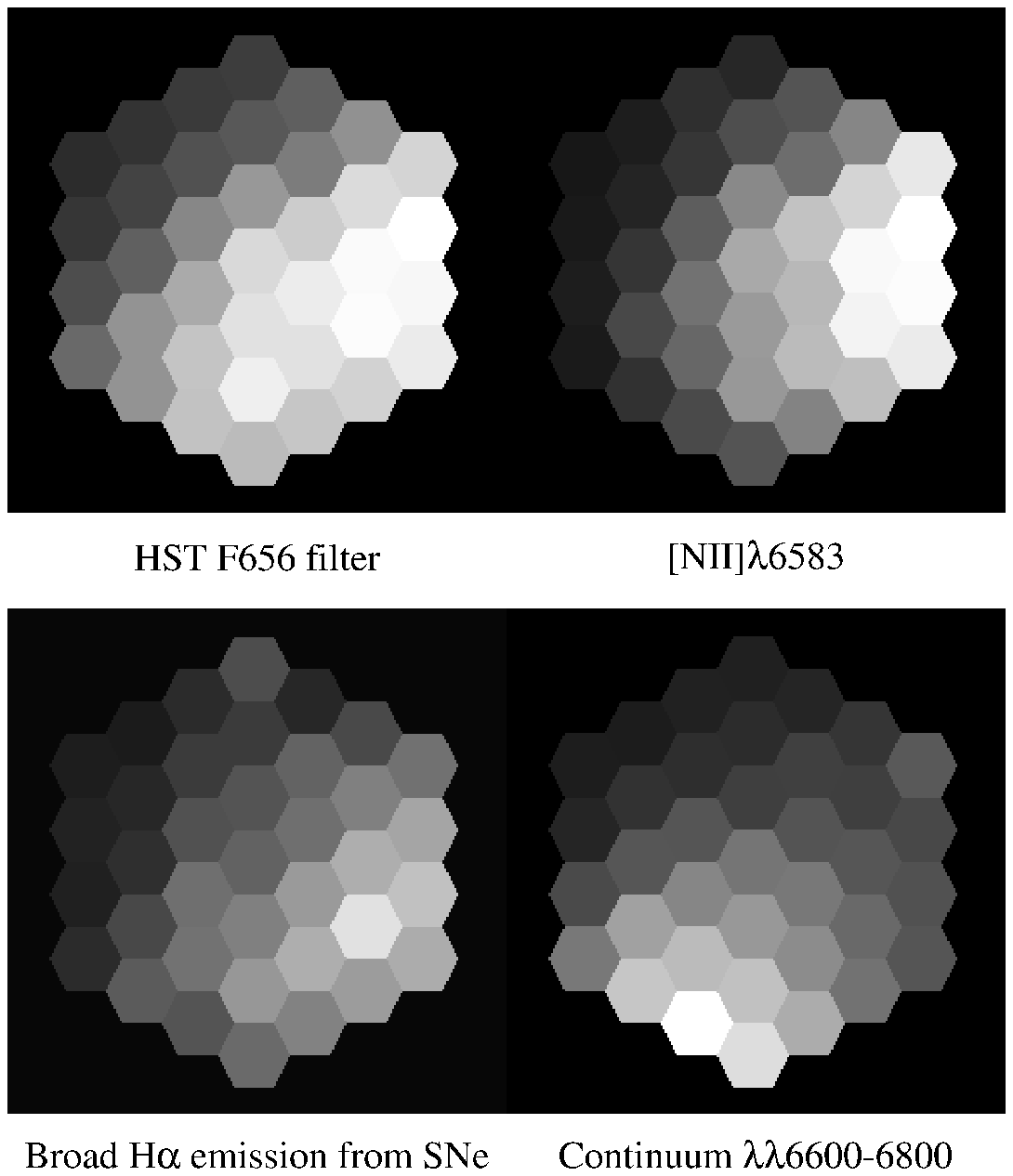}

\figcaption[nfg14.eps]{SPIRAL maps of SN 1987A. All map intensities are normalised. North is top and East is left.\label{fig18}} 

\clearpage
\plotone{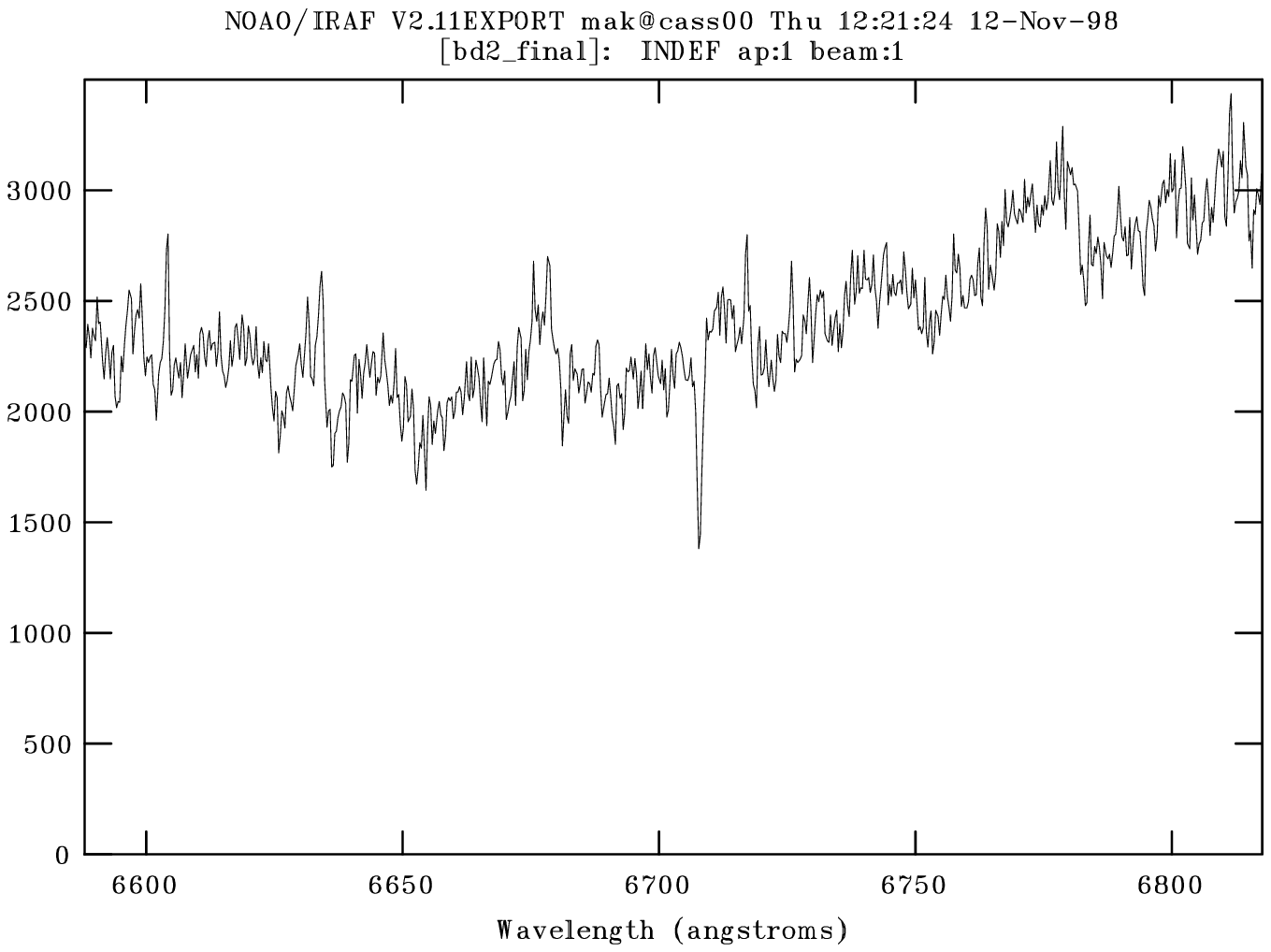}

\figcaption[nfg15.eps]{Spectrum of LP 944-20. The Lithium absorption
line can be seen clearly in the middle of the spectrum. Flux is in
arbitrary units.\label{figbd}}

\end{document}